\documentclass[sigconf,natbib=true,anonymous=false, screen]{acmart}
\usepackage{multirow}
\usepackage{bookmark}
\usepackage{graphicx}
\usepackage{hyperref}
\usepackage{comment}
\usepackage{amsmath,amsfonts}
\usepackage{algorithmic}
\usepackage{textcomp}
\usepackage{xcolor}
\usepackage{float}
\usepackage{subcaption}
\usepackage{enumitem}
\usepackage{bbding}
\usepackage{microtype}
\usepackage{breakcites}
\usepackage{caption}
\AtBeginDocument{%
  }

\renewcommand\footnotetextcopyrightpermission[1]{}

\begin{document}

\title{Revisiting Content-Based Music Recommendation: Efficient Feature Aggregation from Large-Scale Music Models}

\author{Yizhi Zhou}
\email{zhouyz@lamda.nju.edu.cn}
\affiliation{%
  \institution{School of Artificial Intelligence, Nanjing University}
  \institution{National Key Laboratory for Novel Software Technology, Nanjing University}
  \country{NanJing, China}
}

\author{Jia-Qi Yang}
\email{yangjq@lamda.nju.edu.cn}
\affiliation{%
  \institution{School of Artificial Intelligence, Nanjing University}
  \institution{National Key Laboratory for Novel Software Technology, Nanjing University}
  \country{NanJing, China}
}

\author{De-Chuan Zhan}
\email{zhandc@lamda.nju.edu.cn}
\affiliation{%
  \institution{School of Artificial Intelligence, Nanjing University}
  \institution{National Key Laboratory for Novel Software Technology, Nanjing University}
  \country{NanJing, China}
}

\author{Da-Wei Zhou}
\email{zhoudw@lamda.nju.edu.cn}
\affiliation{%
  \institution{School of Artificial Intelligence, Nanjing University}
  \institution{National Key Laboratory for Novel Software Technology, Nanjing University}
  \country{NanJing, China}
}
\authornote{Corresponding author.}

\renewcommand{\shortauthors}{Zhou et al.}

\begin{abstract}
   
 Music Recommendation Systems (MRSs) are a cornerstone of modern streaming platforms. Existing recommendation models, spanning both recall and ranking stages, predominantly rely on collaborative filtering, which fails to exploit the intrinsic characteristics of audio and consequently leads to suboptimal performance, particularly in cold-start scenarios. However, existing music recommendation datasets often lack rich multimodal information, such as raw audio signals and descriptive textual metadata. Moreover, current recommender system evaluation frameworks remain inadequate, as they neither fully leverage multimodal information nor support a diverse range of algorithms, especially multimodal methods. 
To address these limitations, we propose TASTE, a comprehensive dataset and benchmarking framework designed to highlight the role of multimodal information in music recommendation. Our dataset integrates both audio and textual modalities. By leveraging recent large-scale self-supervised music encoders, we demonstrate the substantial value of the extracted audio representations across recommendation tasks, including candidate recall and CTR. In addition, we introduce the \textbf{MuQ-token} method, which enables more efficient integration of multi-layer audio features. This method consistently outperforms other feature integration techniques across various settings. Overall, our results not only validate the effectiveness of content-driven approaches but also provide a highly effective and reusable multimodal foundation for future research. Code is available at \url{https://github.com/zreach/TASTE}
\end{abstract}




\maketitle

\section{Introduction}
With the rise of music streaming, listening habits have shifted from album-based consumption to online playback, fundamentally transforming how music is experienced \cite{bonnin2014automated}. With the rapid growth of music items and users, music recommendation has become increasingly important. Modern deep learning–based recommender systems mainly rely on identifier (ID)-based features and their embeddings to drive learning~\cite{konstas2009social, zheng2024adapting, liu2024learning, zheng2024decentralized}.

  However, the effectiveness of these embeddings for cold-start or long-tail IDs is often limited due to the large amount of interaction data required for training \cite{xu2024cmclrec, lin2024temporally}. Intuitively, the intrinsic modal information of items directly perceived by users can enhance recommender system capabilities and mitigate the above problems. Consequently, researchers are increasingly focused on extracting and leveraging these diverse modalities, including images, text, and audio, and have consistently demonstrated their efficacy \cite{vbpr, lattice, freedom, zhou2023mmrec}. 

Unlike domains such as e-commerce or advertising, music recommendation presents unique characteristics \cite{schedl2018current}.
Music recommendation is complicated by diverse preferences, including genre, style, region, and language. Relying solely on artists or albums is insufficient, as single artists have varied styles and users seek new discoveries. Furthermore, subjective and imprecise user-generated genre tags lack the granularity needed for accurate recommendations. Consequently, by extracting features from the music modality, recommendation systems could more effectively capture user preferences \cite{deldjoo2024content}. While a substantial body of research exists on multimodal recommender systems~\cite{vbpr, mmgcn, lattice, freedom, guo2024lgmrec}, dedicated exploration within the specific context of music remains limited. Although numerous interaction datasets exist within the music domain, limited accessibility of music content poses a significant challenge
to research on content-based recommendation.

To address this gap and support our study, we use datasets Music4All \cite{music4all}, LFM-1b \cite{lfm1b}, and LFM-2b \cite{schedl2022lfm2b}. Based on this setup, we further augment LFM-1b and LFM-2b  by extracting audio and textual representations for a subset of items, resulting in two distinct augmented datasets. To better leverage the audio information of music, we observe that recent advances in music representation learning have led to the development of powerful music models such as MERT \cite{li2023mert}, MusicFM \cite{musicfm}, and MuQ \cite{muq}. These encoders have demonstrated strong performance across Music Information Retrieval (MIR) tasks.
Moreover, these models are trained in a self-supervised learning (SSL) manner, enabling them to learn representations that are less influenced by biases introduced by supervised labels and that better preserve intrinsic musical characteristics. Motivated by these advances, we leverage recent large-scale self-supervised music encoders to process and encode the collected audio data. Furthermore, we propose a novel method 
\textbf{MuQ-token} for efficiently integrating audio features into baseline recommendation models.

\begin{figure*}[t]
    \centering
    \includegraphics[width=\linewidth]{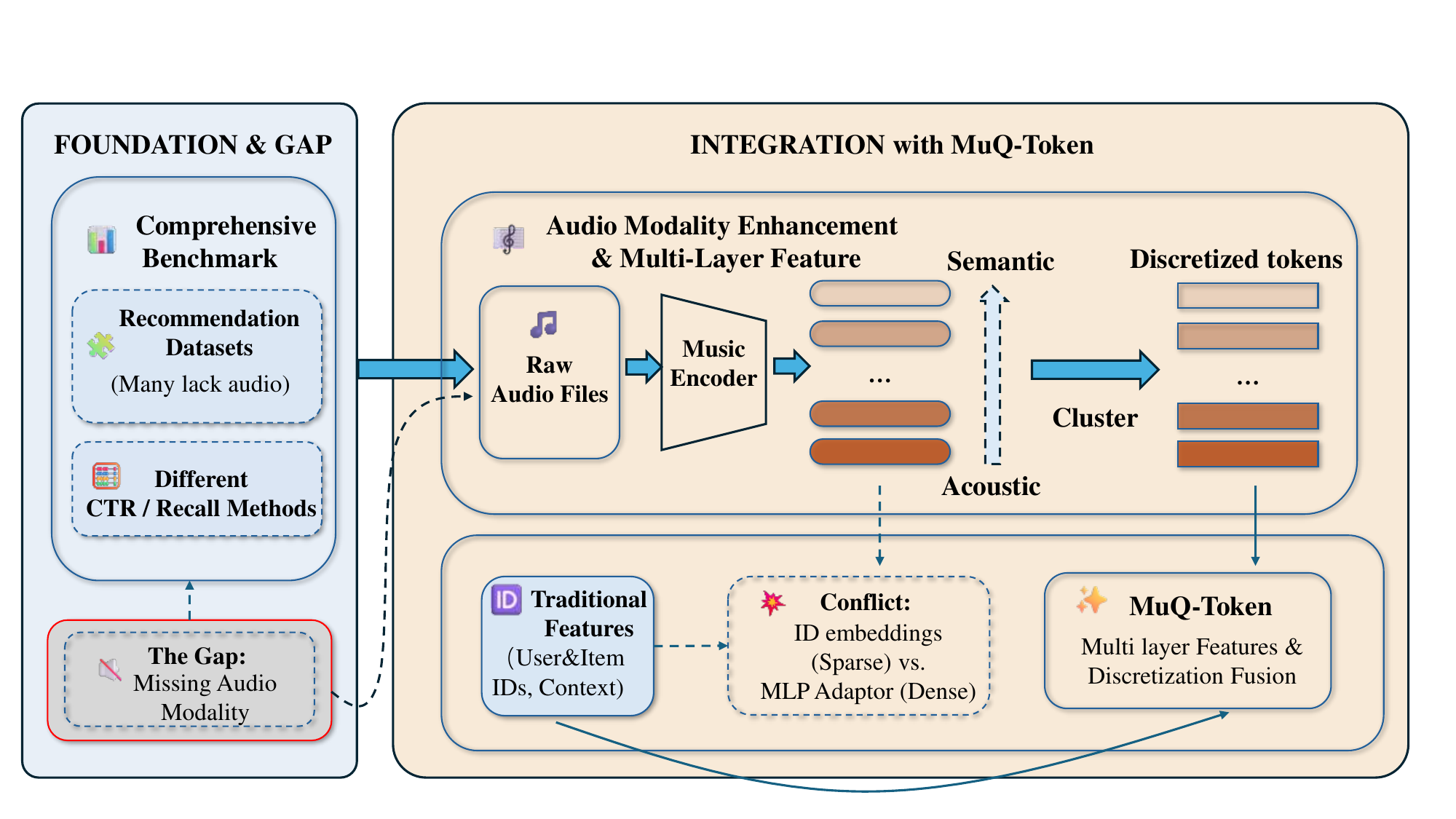}
    \caption{The overall framework of our proposed music recommendation system. The workflow establishes a comprehensive benchmark that integrates multiple datasets and recommendation methods for CTR and recall, and highlights the importance of audio modalities in building effective music recommendation systems. We then collect raw audio data and use music representation models to extract readily usable multi-layer representations to augment the datasets. To address the optimization conflict caused by directly feeding dense features through an MLP adaptor alongside ID embeddings, we introduce a feature discretization module, MuQ-token. Finally, the discretized audio tokens are combined with traditional features, resulting in consistent performance improvements across different methods.}
    \label{fig:framework}
\end{figure*}

While several open-source recommendation benchmarks exist~\cite{recbole[1.0], recbole[1.2.1], recbole[2.0], zhou2023mmrec, DBLP:conf/cikm/ZhuLYZH21, DBLP:conf/sigir/ZhuDSMLCXZ22}, none of them provides a comprehensive framework tailored for multimodal music recommendation.
They typically lack coverage of either click-through rate (CTR) or recall tasks, or do not support multimodal recommendation methods in a unified manner.
More importantly, these benchmarks do not provide readily usable audio or textual modality features, requiring researchers to construct complex data pipelines before meaningful experimentation can begin.
A detailed comparison between existing frameworks and our system is summarized in Table~\ref{tab:benchs-compare}.
As discussed above, leveraging rich audio and text information is essential for advancing music recommendation research. Therefore, we introduce an open-source multimodal music recommender system library called \textbf{TASTE} (con\textbf{T}ent-b\textbf{A}sed mu\textbf{S}ic recommenda\textbf{T}ion b\textbf{E}n\-chmark).  Our comprehensive framework achieves the following key advancements:

\begin{itemize}
    \item \textbf{A large-scale multimodal music recommendation ben\-chmark.} We build a comprehensive multimodal benchmark by augmenting three widely used real-world music interaction datasets (Music4All, LFM-1b, and LFM-2b) with high-quality audio and textual representations. This results in new large-scale audio-enhanced datasets and a unified evaluation suite that supports both CTR prediction and Top-K recommendation. To the best of our knowledge, this is the first benchmark that provides ready-to-use multi-layer, information-rich audio embeddings from SSL music foundation models for music recommendation. %

    \item \textbf{A unified evaluation framework for multimodal music recommendation.} Based on the constructed datasets, we establish an open-source evaluation framework (TASTE) that supports a wide range of recommendation paradigms, including collaborative filtering, context-aware models, and multimodal methods. The framework enables systematic comparisons across CTR and recall tasks, making it easier to study how different modalities and fusion strategies affect recommendation performance, especially under cold-start and long-tail settings.
    
    \item \textbf{An efficient multi-layer audio feature integration me\-thod.} On top of this benchmark, we propose \textbf{MuQ-token}, a novel discretization-based method for integrating multi-layer audio representations into recommender systems. By transforming continuous features from different MuQ layers into semantic tokens, MuQ-token resolves the training conflict between sparse ID embeddings and dense neural features, while effectively exploiting complementary information across layers. Extensive experiments show that MuQ-token consistently outperforms conventional fusion strategies across datasets and models.
    
\end{itemize}

\section{TASTE}

In this section, we formally introduce our proposed dataset and benchmark TASTE. 

\subsection{Overall Framework}

\begin{table*}[!ht]
    \centering
    \caption{Comparison among benchmarks.}
    \begin{tabular}{lcccc}
    \hline
        Benchmark name & CTR task & Recall task & Multimodal Methods & Extracted Multimodal Features \\ \hline
        MMRec \cite{zhou2023mmrec} & $\times$ & $\checkmark$ & $\checkmark$  & $\times$ \\ 
        Recbole \cite{recbole[1.0], recbole[1.2.1], recbole[2.0]}& $\checkmark$ & $\checkmark$ & $\times$ & $\times$ \\ 
        FuxiCTR \cite{DBLP:conf/cikm/ZhuLYZH21, DBLP:conf/sigir/ZhuDSMLCXZ22} & $\checkmark$ & $\times$ & $\times$ & $\times$ \\
        TASTE (ours) & $\checkmark$ & $\checkmark$ & $\checkmark$ & $\checkmark$ \\ \hline
    \end{tabular}
    
    \label{tab:benchs-compare}
\end{table*}

As illustrated in Figure \ref{fig:framework}, TASTE is a multimodal benchmarking and evaluation framework that integrates multiple datasets with a diverse set of CTR and recall algorithms.

First, TASTE starts from real-world music interaction datasets, which typically provide user–item interaction data along with limited metadata. However, such datasets often lack audio content that can be directly leveraged for multimodal recommendation; even when audio-related information is available, it is usually restricted to basic features such as Mel-Frequency Cepstral Coefficients (MFCCs). To address this limitation, we collect and align raw audio signals for a subset of items and encode them using large self-supervised music foundation models. This process produces multi-layer audio representations, where different layers capture distinct musical properties.

For feature integration, we initially adopt a straightforward approach by using a trainable multi-layer perceptron (MLP) adaptor to project the audio features into the same dimensional space as ID-based embeddings and concatenate them. However, jointly training continuous audio representations with discrete ID embeddings can lead to performance degradation, as discussed in \cite{taobao}. To overcome this issue, we introduce MuQ-token, a discretization-based feature integration method that converts layer-wise audio representations into discrete tokens, effectively mitigating the aforementioned problem. Finally, the discretized audio tokens are fused with conventional recommendation features and evaluated using a wide range of CTR and recall models within the TASTE framework. This unified pipeline enables fair and systematic assessment of multimodal contributions, consistently improving performance on both CTR prediction and Top-K recall tasks. Moreover, all extracted multimodal features are released in a plug-and-play manner to support further research.

\subsection{Data Preparation}
\subsubsection{Datasets}
Our study is based on three widely used real-world music recommendation datasets, which provide large-scale user--item interaction logs together with rich side information.

\textbf{Music4All}~\cite{music4all} is a music database that contains metadata, tags, genre annotations, 30-second audio clips, lyrics, and user interaction records. The interaction data span from January 1st, 2019, 00:00 to March 20th, 2019, 23:59.

\textbf{LFM-1b}~\cite{lfm1b} is a large-scale music listening dataset containing artists, albums, tracks, and users, together with individual listening events from January 2013 to August 2014. It also provides diverse metadata, including genre labels, which makes it suitable for content-based and multi-label learning tasks.

\textbf{LFM-2b}~\cite{schedl2022lfm2b} contains over two billion listening events from more than 120,000 Last.fm users, covering a time span from February 2005 to March 2020. The dataset includes more than 50 million tracks and 5 million artists, together with rich metadata such as genres, styles, user demographics, and vector embeddings of lyrics.

\begin{table}[!ht]
    \centering
    \caption{Datasets' Statistical Properties (The "inter" column counts the total number of unique user-item interaction pairs. Duplicate interactions are counted only once.)}
    \begin{tabular}{lcccc}
    \hline
        Dataset & user & item & inter  & audio features \\ \hline
        music4all & 15,602  & 109,269 & 2,597,382 & 30s-clips \\
        lfm-1b & 120,322 & 31,413,999 & 249,995,221 & $\times$ \\ 
        lfm-1b-taste & 113,891 & 85,711 & 4,976,486 & $\checkmark$ \\
        lfm-2b & 120,322 & 50,813,373 & 519,293,333 & $\times$ \\ 
        lfm-2b-taste & 116,543 & 95,059 & 10,231,198 & $\checkmark$ \\
        \hline
    \end{tabular}
    
    \label{tab:statics}
\end{table}

To incorporate audio modality into these interaction datasets, we collect and align audio signals for a subset of items in each dataset and extract their representations using multiple state-of-the-art music feature extractors. This process yields three audio-enhanced datasets that preserve the original interaction structures while providing rich acoustic information for each item. These constructed datasets are used throughout our study, and their overall statistics are summarized in Table~\ref{tab:statics}.

Unlike many explicit-feedback benchmarks, music listening data does not provide ground-truth ``positive'' or ``negative'' labels for user--item interactions. To enable CTR modeling, we therefore convert the implicit feedback into binary supervision. For each user--item pair, we compute the total number of interactions, denoted as $count$. Pairs with $count$ greater than or equal to a predefined threshold are treated as positive samples, while those below the threshold are regarded as negative. We set the threshold to 2 for all datasets.

Finally, following standard practice in recommendation benchmarks, we apply a 5-core filtering on users and items, retaining only those that appear in at least five interactions, in order to ensure data quality and robustness.
    
\subsection{Multimodal Features} \label{sec:encoders}

\subsubsection{Audio Features} The foundation models are pre-trained on a large amount of unlabeled data to learn high-dimensional representations of audio. It aims to compress and denoise long raw audio signals, retaining as much information in signals as possible. The training objective is to predict tokens that have been randomly masked within a sequence. similar to BERT \cite{devlin2019bert}. In audio processing, authors used a variety of feature extractors to generate the tokens for training. MuQ use Mel-RVQ.

\textbf{MuQ}\cite{muq} It is a large music foundation model pre-trained via Self-Supervised Learning (SSL). So far, it has demonstrated state-of-the-art results across all music information retrieval tasks.

\textbf{MuQ-Mulan} It is an alternative version of MuQ, fine-tuned on text-audio pairs using a CLIP-like contrastive learning approach~\cite{clip}. It supports both Chinese and English inputs.

\textbf{CLAP}\cite{CLAP2022, CLAP2023}. It is a model that learns acoustic concepts from natural language supervision and enables “Zero-Shot” inference. The model has been extensively evaluated in 26 audio downstream tasks, achieving SoTA in several of them, including classification, retrieval, and captioning. It has been previously used and proven effective by works in the music recommendation domain~\cite{salganik2024larp,pan2025bridging}.

\textbf{MFCCs} Mel-Frequency Cepstral Coefficients are a widely used audio feature that mimics how humans perceive sound~\cite{deepmusic}. They transform complex audio signals into a compact representation, primarily capturing the timbre or ``color'' of a sound, making them ideal for tasks like speech and music analysis. We use this feature as the baseline for our audio features.

\subsubsection{Features Preprocessing}

MuQ extracts features of shape\\ $(L, T, H)$ with a 75 ms frame rate. We apply average pooling over the temporal dimension to obtain $(L, H)$ representations, where the number of layers $L$ is discussed later. In our setup, $H=1024$ for MuQ. For MFCCs, we similarly apply temporal average pooling to produce an $(H)$-dimensional vector with $H=80$. CLAP and MuQ-Mulan directly output fixed-dimensional $(H)$ representations and thus require no additional pooling.

\subsubsection{Text Features} Regarding text features, we utilized the categorical text of all musical pieces and encoded them with MuQ-Mulan's text encoder. The features are all of shape (d), $d=512$. When audio is unlabeled, we substitute it with an all-zero vector.

\subsection{MuQ-token} \label{section:token}

State-of-the-art music foundation models such as MuQ produce multi-layer audio representations, where different layers capture complementary musical characteristics. These layer-wise features serve as the input to our feature integration module.

A natural way to integrate such multi-layer representations is to either (i) use each layer independently as input features, or (ii) aggregate all layers into a single vector by averaging or weighted pooling. We evaluate these strategies in our experiments and select the best-performing one on the validation set. After aggregation, the resulting feature vector is projected by a multi-layer perceptron (MLP) to match the dimensionality of other embeddings before being fed into CTR or recall models.

However, these continuous fusion strategies suffer from two fundamental limitations. First, dense neural features often interfere with sparse ID embeddings during joint optimization, leading to unstable training and suboptimal convergence. Second, pooling-based aggregation tends to blur the complementary information encoded across different layers, making it difficult for the model to exploit fine-grained distinctions that are distributed over the depth of the encoder. \cite{zhou2025layer}

To address these issues, we propose \textbf{MuQ-token}, a discretization-based integration method that converts multi-layer audio representations into layer-wise tokens compatible with standard recommendation architectures. Specifically, for each MuQ layer 
l, we collect all feature vectors from that layer across the dataset and apply K-means clustering. Each audio track is then assigned a cluster ID for each layer, yielding a sequence of discrete tokens—one per layer—which are treated as categorical features and embedded in the same way as user IDs, item IDs, or genre tags.

\begin{figure}[H]
    \centering
    
    \includegraphics[width=\linewidth]{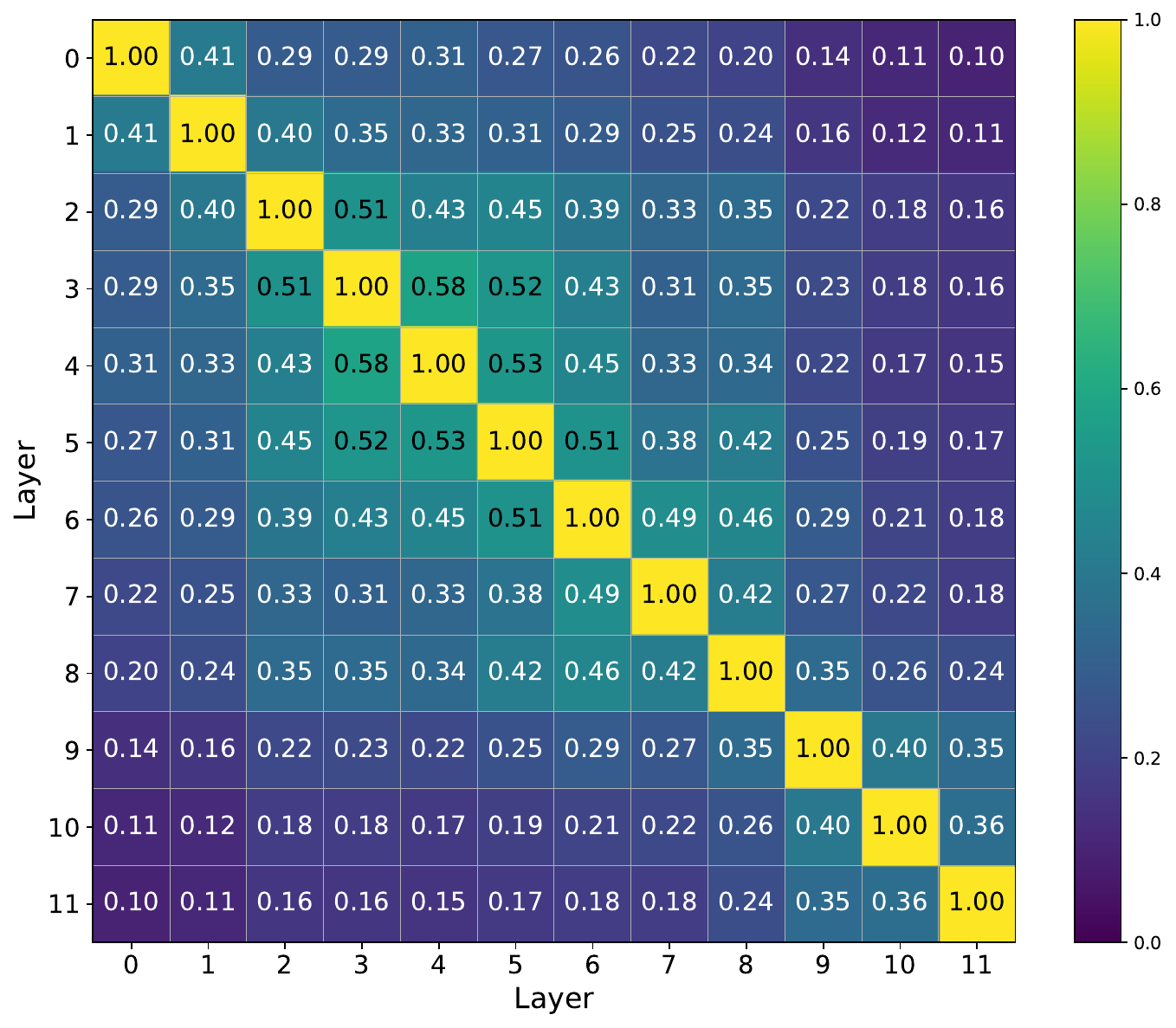}
    \caption{Similarity of Clustering Results Across Different MuQ Output Feature Layers in the Music4all Dataset. }
    \label{fig:ari_32}
\end{figure}

By design, this tokenization avoids the optimization conflict between continuous audio features and sparse ID embeddings, while allowing each layer to contribute independently to the final prediction. Importantly, MuQ-token preserves the diversity of representations across layers instead of collapsing them into a single pooled vector. To verify that different layers indeed encode complementary information, we measure the similarity between clustering results using the Adjusted Rand Index (ARI). As shown in Figure \ref{fig:ari_32}, adjacent layers exhibit higher ARI scores, indicating partial overlap, while substantial differences remain across distant layers. This shows that the layers encode distinct, non-redundant structures
rather than collapsing into a single representation space,
supporting our decision to jointly use tokens from all layers
instead of relying on a single aggregated representation.

Through this discretization-based multi-layer integration, MuQ-token effectively captures diverse musical characteristics encoded at different depths of the MuQ model, leading to consistent improvements over conventional fusion strategies in both CTR and recall tasks.

\section{Experiment}

\subsection{Problem Formulation}
\subsubsection{Top-k Recommendation}

Formally, let $\mathcal{U}$ be the set of all users and $\mathcal{I}$ be the set of all items. We are given a historical interaction dataset $\mathcal{D} = \{ (u, i, t) \}$ where $u \in \mathcal{U}$ is a user, $i \in \mathcal{I}$ is an item, and $t$ represents the timestamp of the interaction. This dataset implicitly captures user preferences and behaviors.

Our objective is to learn a function $f: \mathcal{U} \times \mathcal{I} \rightarrow \mathbb{R}$ that estimates the relevance or preference of user $u$ for item $i$. Based on this estimated relevance score, for any given user $u$, we aim to generate a ranked list of items $\mathcal{L}_u = [i_1, i_2, \dots, i_K]$ from $\mathcal{I} \setminus \mathcal{I}_u^{hist}$ (where $\mathcal{I}_u^{hist}$ is the set of items user $u$ has already interacted with), such that items with higher predicted relevance appear earlier in the list. The ultimate goal is to provide users with items they are most likely to engage with or find valuable, thereby enhancing their experience with the system. The evaluation metrics are as follows:

\textbf{Recall@K}
Measures the proportion of relevant items retrieved within the top-$K$ recommendations.
$$\text{Recall@K} = \frac{|R_u \cap P_u|}{|R_u|}$$
where $R_u$ is the set of relevant items and $P_u$ is the set of top-$K$ recommended items for user $u$.

\textbf{Mean Reciprocal Rank (MRR)}
Evaluates the position of the first relevant item in a ranked list.
$$\text{MRR} = \frac{1}{|Q|} \sum_{q=1}^{|Q|} \frac{1}{\text{rank}_q}$$
where $\text{rank}_q$ is the rank of the first relevant item for query $q$.

\textbf{Normalized Discounted Cumulative Gain (NDCG@K)}
Considers both relevance and position in the ranked list, penalizing relevant items at lower ranks.
$$\text{NDCG@K} = \frac{\text{DCG}_K}{\text{IDCG}_K}$$
where $\text{DCG}_K = \sum_{i=1}^{K} \frac{2^{\text{rel}_i} - 1}{\log_2(i + 1)}$ and $\text{IDCG}_K$ is the maximum possible DCG.


\textbf{Precision@K}
Measures the proportion of recommended items that are actually relevant among the top $K$.
$$\text{Precision@K} = \frac{|R_u \cap P_u|}{|P_u|}$$
where $R_u$ is the set of relevant items and $P_u$ is the set of top-$K$ recommended items for user $u$.

\subsubsection{CTR Prediction}
Click-Through Rate (CTR) prediction estimates the probability that a user will click on an item. It is a core task in recommendations and online advertising, aiming to rank items to maximize user engagement.

Given a user $u$, an item $i$, and their features, we want to predict the probability $P(click | u, i, \text{context})$. This is a binary classification problem, learning a function\\  $f: (\text{features}_u, \text{features}_i, \text{features}_{\text{context}}) \rightarrow [0, 1]$ from historical positive/negative data. The evaluation metrics are as follows:


\textbf{Log Loss}: Measures prediction accuracy.
    $$\text{LogLoss} = -\frac{1}{N} \sum_{i=1}^{N} [y_i \log(p_i) + (1 - y_i) \log(1 - p_i)]$$


\textbf{Area Under the ROC Curve (AUC)}: Measures the model’s discriminative ability between clicked and non-clicked instances, and is robust to class imbalance.

\subsection{Baselines}

To comprehensively evaluate the effectiveness of multimodal audio features, we benchmark a broad range of representative models for both CTR prediction and recall tasks.

\paragraph{CTR Models.}
We consider three categories of CTR predictors:
(i) \emph{linear models}, represented by Logistic Regression (LR)~\cite{LR};
(ii) \emph{second-order factorization models}, including
FM~\cite{FM}, FFM~\cite{FFM}, and AFM~\cite{AFM};
and (iii) \emph{deep feature interaction models}, including\\
Wide\&Deep~\cite{wide}, DeepFM~\cite{guo2017deepfm},
xDeepFM~\cite{lian2018xdeepfm}, NFM~\cite{NFM},
DCN~\cite{DCN}, DCNv2~\cite{DCNv2},
MaskNet~\cite{wang2021masknet},
FinalMLP~\cite{mao2023finalmlp},
EulerNet~\cite{tian2023eulernet},
WuKong~\cite{zhang2024wukong},
and FiGNN~\cite{li2019fi}.
Together, these models span the most widely adopted architectures
for capturing low- to high-order feature interactions
in industrial CTR systems.

\paragraph{Recall Models.}

For the recall task, we adopt Bayesian Personalized Ranking (\textbf{BPR} \cite{BPR}) as the ID-only baseline. We further include several classical and widely adopted multimodal recall models, including \textbf{VBPR} \cite{vbpr}, \textbf{FREEDOM} \cite{freedom}, and \textbf{LGMRec} \cite{guo2024lgmrec}. These methods enhance the basic collaborative filtering framework by integrating content features through linear fusion mechanisms or graph-based multimodal representation learning, enabling a comprehensive evaluation of the recall methods.

\paragraph{Multimodal Integration.}
For CTR models, we consider three settings:
(a) an ID-only baseline without multimodal features;
(b) concatenating MuQ features from all layers with other input features after passing them through a trainable MLP adaptor; and
(c) using the discretized outputs of MuQ-token, where features from each layer are treated as separate categorical inputs. For the recall task, we adopt the multimodal fusion mechanisms provided by each algorithm.

\subsection{Feature Embedding Methods} \label{sec:feat_emb}
In the CTR task, different types of features are input. Generally, the more features there are, the better the performance. The data types can be categorized into the following types:
\begin{itemize}
    \item \textbf{Discrete Features:} User/item IDs, gender, country, etc., are primarily \textbf{one-hot encoded}. Each category forms a ``field'' with a corresponding vector where one element is ``1'' (hot) and others ``0'' (cold). For multi-valued fields (e.g., music genres), multiple elements can be ``1''. This converts categorical data into a numerical format for machine learning models.

    \item \textbf{Numeric Features:} Age, play count, etc., are continuous. Directly treating them as discrete values or simply concatenating them after normalization (direct inclusion) often leads to performance loss due to poor feature expression. Therefore, we typically use different discretization methods to process these features.
    \begin{itemize}
        \item \textbf{Field embedding} multiplies a learnable vector by the continuous value, but this can limit expressiveness due to linear relationships.
        \item \textbf{Hard discretization} Based on the maximum and minimum values of the feature, set up k buckets with a width of $w=\left(x_{\max }-x_{\min }\right) / k$ and evenly divide them according to a specific criterion. Then, we can assign feature values to the corresponding buckets based on their values with $x^{\prime}=\left(x-x_{\min }\right) / w, \quad x^{\prime} \in[0, k)$
        \item \textbf{Soft discretization} Some works tend to dynamically bucketize the numeric feature \cite{autodis}.
    \end{itemize}

\end{itemize}

These input methods were used for our baseline models. Through experimental trials, we ultimately chose equal-width binning as the discretization method for numerical features.

\subsection{Experimental Settings}

During training, we used the Adam algorithm to optimize the model, with a learning rate of 0.001. To prevent overfitting, we applied early stopping, halting training early when performance on the validation set did not improve over several epochs. The train batch size was set to 2048. We experimented to find the optimal embedding size for each model. Then, to ensure fairness, we used that same embedding size when testing different settings within the same method.
We randomly divided each user's listening history into training, validation, and test sets with an 8:1:1 ratio.

For the CTR task, we experimented with several feature settings:
(1)  IDonly: Using only user and item IDs.
(2)  Categorical: Incorporating user and item IDs along with categorical information.
(3)  Full Feature Set: Employing user IDs, item IDs, categorical, and numerical features. 
These features are then embedded using the method described in subsection \ref{sec:feat_emb} before being fed into the model.

On top of these base configurations, we explored two main approaches for integrating additional modal information:

\begin{enumerate}
    \item \textbf{MuQ}: We added features encoded by MuQ and encoded text features. For non-discretized information, we tested both inputting features from all layers and using mean-pooled features, ultimately selecting the best-performing setup based on the validation set for final testing.
    \item \textbf{MuQ-token}: We integrated MuQ-Token (discretized cluster IDs from multiple MuQ layers as described in \ref{section:token}) and text features. The number of clusters was set to 16, based on multiple experiments.
\end{enumerate}
In the recall task, we employed the feature fusion approaches as introduced in each of their original papers.

\subsection{Main Results} 
The performance comparison for all methods on the three
datasets is summarized in Table \ref{tab:result1} and \ref{tab:result-recall}, from which we have the following key observations:
 \begin{enumerate}
     \item Significant Gains from MultiModal Features: For CTR tasks, incorporating modal features consistently improved metrics across various datasets and models, outperforming models that lacked this information. Even with the simplest model, FM, adding modal data led to enhancements. Similarly, for the recall task, multimodal models surpassed ID-only approaches; even VBPR, which linearly combines modal information with BPR, showed better performance. This suggests that, in practice, integrating modal information generally enhances system capabilities.
     \item Significant performance gains were observed with the inclusion of clusterID, benefiting nearly all models. This demonstrates our successful utilization of multi-layer information. In prior research, each layer of an unsupervised music representation model was understood to represent different levels of information, ranging from acoustic to semantic. Different users, however, may care about distinct aspects of music preferences. For example, some user groups might focus on low versus high frequencies, or prominent guitars versus bass, while others might be more interested in the emotional expression of the music, and so on. Our approach, through an efficient and straightforward method, effectively leverages information from every layer, leading to overall improvements.
     \item The benefits of adding modal features depend on the model's nature. We observed that first-order models (LR) gain no advantage from them, and even basic second-order interaction models (FM, FFM) see very modest increases. Interestingly, even among second-order interaction models, the AFM model—which adds an attention mechanism to FM—achieved notable improvements with multimodal features. This suggests that users are often attracted to distinct aspects of musical data, and the attention mechanism effectively helps the model identify these varied preferences.
 \end{enumerate} 

\renewcommand{\arraystretch}{1.05}
\begin{table*}[!ht]
\caption{Main Results of AUC(\%) in the CTR task $\uparrow$. ``MuQ-token'' denotes the feature representation approach that utilizes cluster IDs. In each dataset, the best result for each method is underlined, and the best overall result among all methods is shown in bold.}
    \begin{tabular}{cccc|ccc|ccc}
    \hline \multirow{2}{*}{ Model } & \multicolumn{3}{c}{ m4a } & \multicolumn{3}{c}{ lfm-2b-taste } & \multicolumn{3}{c}{ lfm-1b-taste } \\
    \cline { 2 - 10 } & w/o multimodal  &  MuQ & MuQ-token & w/o multimodal  & MuQ & MuQ-token & w/o multimodal  & MuQ & MuQ-token \\
    \hline
LR       & \underline{78.31} & 78.25 & 78.26 & \underline{81.27} & 81.21 & 81.21 & \underline{80.78} & 80.73 & 80.74 \\
\hline
FM       & 78.83 & \underline{80.92} & 80.22 & 85.17 & \underline{85.25} & 84.82 & 84.48 & \underline{84.67} & 84.17 \\
FFM      & 79.25 & 79.33 & \underline{80.05} & \underline{84.06} & 84.04 & 83.97 & 83.34 & 83.45 & \underline{83.83} \\
AFM      & 79.85 & 81.48 & \underline{82.80} & 85.12 & 85.47 & \underline{85.83} & 84.90 & 84.78 & \underline{85.17} \\
\hline
FiGNN    & 79.69 & 81.50 & \underline{81.99} & 84.31 & 85.08 & \underline{85.74} & 83.87 & 84.24 & \underline{84.84} \\
WideDeep & 79.94 & 81.53 & \underline{81.56} & 84.02 & 84.84 & \underline{86.00} & 84.58 & 84.78 & \underline{85.45} \\ 
DeepFM   & 79.79 & \underline{81.78} & 81.26 & 83.81 & \underline{84.06} & 83.86 & \underline{83.43} & 83.16 & 83.37 \\
NFM      & 79.77 & 82.54 & \underline{82.64} & 83.32 & 85.85 & \underline{86.02} & 82.24 & 83.87 & \underline{85.23} \\ 
xDeepFM  & 79.30 & 82.21 & \underline{82.29} & 82.03 & 85.52 & \underline{85.84} & 83.43 & 83.79 & \underline{85.18} \\
DCN      & 80.39 & 82.14 & \underline{82.41} & 85.22 & 85.90 & \underline{86.62} & 85.37 & 85.23 & \underline{86.18} \\ 
DCNv2    & 80.03 & 82.36 & \underline{82.97} & 85.60 & 86.58 & \textbf{86.89} & 85.36 & 85.55 & \textbf{86.21} \\ 
MaskNet  & 79.88 & 82.92 & \textbf{83.22} & 84.78 & 86.57 & \underline{86.68} & 83.85 & 85.23 & \underline{85.93} \\
FinalMLP & 79.88 & \underline{81.69} & 81.45 & 85.79 & 86.29 & \underline{86.53} & 85.05 & 85.05 & \underline{85.95} \\
EulerNet & 80.45 & 81.35 & \underline{82.57} & 85.01 & 85.64 & \underline{86.53} & 85.40 & 85.57 & \underline{85.96} \\
WuKong   & 79.10 & 82.34 & \underline{82.54} & 83.64 & 84.57 & \underline{84.78} & 82.72 & 84.11 & \underline{84.95} \\
\hline
\multicolumn{10}{c}{ ID+Categories } \\
\hline
LR       & \underline{78.30} & 78.24 & 78.26 & \underline{81.27} & 81.24 & 81.20 & \underline{80.78} & 80.73 & 80.74 \\
\hline
FM       & 79.65 & \underline{80.85} & 80.23 & 85.26 & \underline{85.33} & 84.82 & \underline{84.61} & 84.55 & 84.22 \\
FFM      & 79.49 & 79.55 & \underline{80.49} & 83.50 & 84.22 & \underline{84.41} & \underline{83.98} & 83.95 & 83.98 \\ 
AFM      & 80.10 & 81.96 & \underline{82.86} & 85.41 & 85.89 & \underline{86.26} & 84.62 & 85.23 & \underline{85.49} \\
\hline
FiGNN    & 80.45 & 81.96 & \underline{81.99} & 84.21 & \underline{85.94} & 85.91 & 84.41 & 84.89 & \underline{85.16} \\
WideDeep & 80.56 & 81.86 & \underline{81.86} & 84.50 & 85.87 & \underline{86.17} & 84.88 & 84.99 & \underline{85.57} \\ 
DeepFM   & 80.46 & \underline{81.86} & 81.46 & 84.11 & \underline{84.29} & 83.85 & 83.37 & \underline{83.68} & 83.56 \\
NFM      & 80.72 & 82.60 & \underline{82.79} & 85.08 & 86.19 & \underline{86.38} & 83.82 & 85.48 & \underline{85.55} \\ 
xDeepFM  & 79.90 & 82.73 & \underline{82.83} & 85.21 & 85.60 & \underline{86.12} & 84.56 & \underline{85.25} & 85.24 \\
DCN      & 80.93 & 82.34 & \underline{82.77} & 85.87 & 86.64 & \underline{86.87} & 85.87 & 85.93 & \underline{86.41} \\ 
DCNv2    & 81.04 & 82.87 & \underline{83.10} & 86.54 & 86.91 & \textbf{87.10} & 85.81 & 86.00 & \textbf{86.47} \\ 
MaskNet  & 80.82 & 83.28 & \textbf{83.66} & 85.86 & 87.00 & \underline{87.07} & 85.05 & \underline{86.07} & 86.29 \\
FinalMLP & 80.41 & 82.07 & \underline{81.83} & 85.55 & 86.53 & \underline{86.82} & 85.58 & 85.67 & \underline{86.15} \\
EulerNet & 81.39 & 82.08 & \underline{82.61} & 85.64 & 86.55 & \underline{86.93} & 85.75 & 85.73 & \underline{86.25} \\
WuKong   & 80.16 & 82.41 & \underline{82.70} & 84.66 & 85.02 & \underline{85.31} & 82.97 & 83.01 & \underline{83.22} \\
\hline
\multicolumn{10}{c}{ ID+Categories+Numeric} \\
\hline
LR & \underline{78.34} & 78.30 & 78.34 & \underline{81.27} & 81.24 & 81.20 & \underline{80.70} & 80.70 & 80.66 \\
\hline
FM & 79.84 & \underline{80.58} & 80.38 & 85.26 & \underline{85.33} & 84.82 & \underline{84.34} & 84.29 & 84.11 \\ 
FFM & 80.34 & 80.35 & \underline{80.92} & 84.22 & 84.25 & \underline{84.27} & \underline{85.53} & 85.48 & 85.15 \\ 
AFM & 80.37 & 82.28 & \underline{82.96} & 85.41 & 85.89 & \underline{86.26} & 86.47 & 86.45 & \underline{86.75} \\
\hline
FiGNN & 80.92 & \underline{82.42} & 81.86 & 84.96 & \underline{85.94} & 85.91 & 85.49 & 85.72 & \underline{85.91} \\ 
WideDeep & 81.26 & 81.83 & \underline{82.09} & 85.38 & 85.63 & \underline{85.92} & 85.58 & 85.62 & \underline{85.97} \\ 
DeepFM & 80.94 & \underline{81.86} & 81.39 & 84.11 & \underline{84.24} & 83.93 & 83.72 & \underline{83.78} & 83.40 \\
NFM & 81.25 & 82.31 & \underline{82.98} & 85.10 & 86.19 & \underline{86.42} & 85.78 & 86.00 & \underline{86.50} \\ 
xDeepFM & 80.51 & 81.83 & \underline{82.41} & 85.21 & 86.07 & \underline{86.13} & 85.70 & 85.87 & \underline{86.29} \\
DCN & 81.41 & 82.64 & \underline{82.91} & 86.43 & 86.73 & \underline{86.88} & 86.75 & 86.76 & \underline{86.92} \\ 
DCNv2 & 81.52 & 82.62 & \underline{83.32} & 86.47 & 87.01 & \underline{87.03} & 86.81 & 86.81 & \underline{87.15} \\ 
MaskNet & 81.63 & 83.17 & \textbf{83.61} & 85.94 & 87.01 & \textbf{87.11} & 86.98 & 87.00 & \textbf{87.19} \\
FinalMLP & 81.17 & 82.15 & \underline{82.06} & 86.25 & 86.55 & \underline{86.70} & 86.40 & 86.36 & \underline{86.66} \\
EulerNet & 81.43 & 82.54 & \underline{82.66} & 86.44 & 86.56 & \underline{86.68} & 86.83 & 86.88 & \underline{86.93} \\
WuKong & 81.22 & 82.13 & \underline{82.45} & 85.10 & 85.36 & \underline{85.44} & 85.33 & 85.72 & \underline{85.94} \\
\hline

    \end{tabular}
    
    \label{tab:result1}
\end{table*}

\begin{table*}[!ht]
\caption{Main Results in the recall Task(\%) $\uparrow$ R, M, N, and P stand for Recall, MRR, NDCG, and Precision, respectively. }
    \begin{tabular}{ccccc|cccc|cccccc}
    \hline \multirow{2}{*}{ Model } & \multicolumn{4}{c}{ m4a } & \multicolumn{4}{c}{ lfm-2b-taste } & \multicolumn{4}{c}{ lfm-1b-taste } \\
    \cline { 2 - 13 } & R@10 & M@10 & N@10 & P@10 & R@10 & M@10 & N@10 & P@10 & R@10 & M@10 & N@10 & P@10 \\
    \hline
        BPR & 0.0391 & 0.1898 & 0.069 & 0.0523 & 0.0464 & 0.0845 & 0.0441 & 0.0282 & 0.0533 & 0.0771 & 0.0437 & 0.0250 \\ 
        VBPR & 0.0530 & 0.2178 & 0.0868 & 0.0702 & 0.0528 & 0.0911 & 0.0486 & 0.0302 & 0.0729 & 0.1123 & 0.0629 & 0.0361 \\ 
        FREEDOM & 0.0541  & \textbf{0.2456} & 0.0909 & 0.0760 & 0.0571  &  0.1088 &  0.0559 & 0.0354 & 0.0862 & 0.1214 & 0.0711 & 0.0378 \\
        LGMRec & \textbf{0.0570} & 0.2311 & \textbf{0.0920}   &\textbf{0.0791} & \textbf{0.0793} & \textbf{0.1388} & \textbf{0.0794} & \textbf{0.0489} & \textbf{0.0954} & \textbf{0.1311} & \textbf{0.0812} & \textbf{0.0412} \\ 

    \hline
    \end{tabular}
    \label{tab:result-recall}
\end{table*}

\begin{figure}[H]
    \centering
    \includegraphics[width=\linewidth]{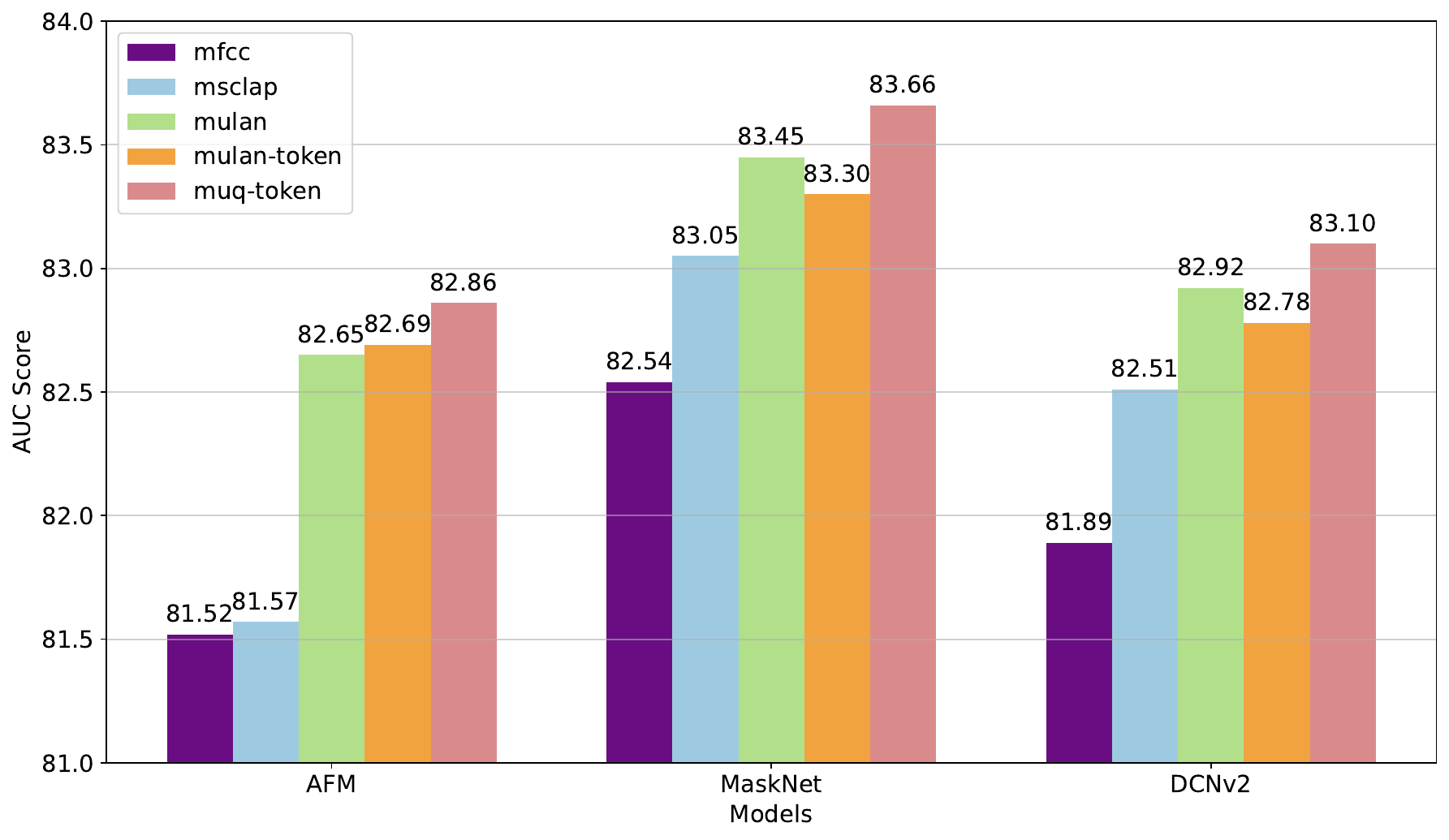}
    \caption{Model Performance Comparison Across Different Audio Features. Here, “mulan-token” denotes the setting where features extracted by MuQ-Mulan are discretized and used as model inputs.}
    \label{fig:encoders}
\end{figure}

\subsection{Ablation Study on Music Encoders}

Is MuQ-token the best choice? To validate this, we experimented with several models on the Music4all dataset using the ``ID + Categories + Numeric with Multimodal features'' configuration. The results, as shown in Figure  \ref{fig:encoders}, indicate that MuQ-token outperforms other audio fusion methods.

\subsection{Diversity of Recommendation} \label{sec:diversity}

We set up an experiment to see if multimodal information boosts the diversity of recommendations. For the music4all dataset, we tested two setups: one with all metadata and another with metadata plus modal information. We chose this dataset because it contains extensive metadata for many items, including language, tempo, key, energy... Without loss of generality, we used the DCNv2 model. 

\begin{figure}[H]
    \centering
    \includegraphics[width=\linewidth]{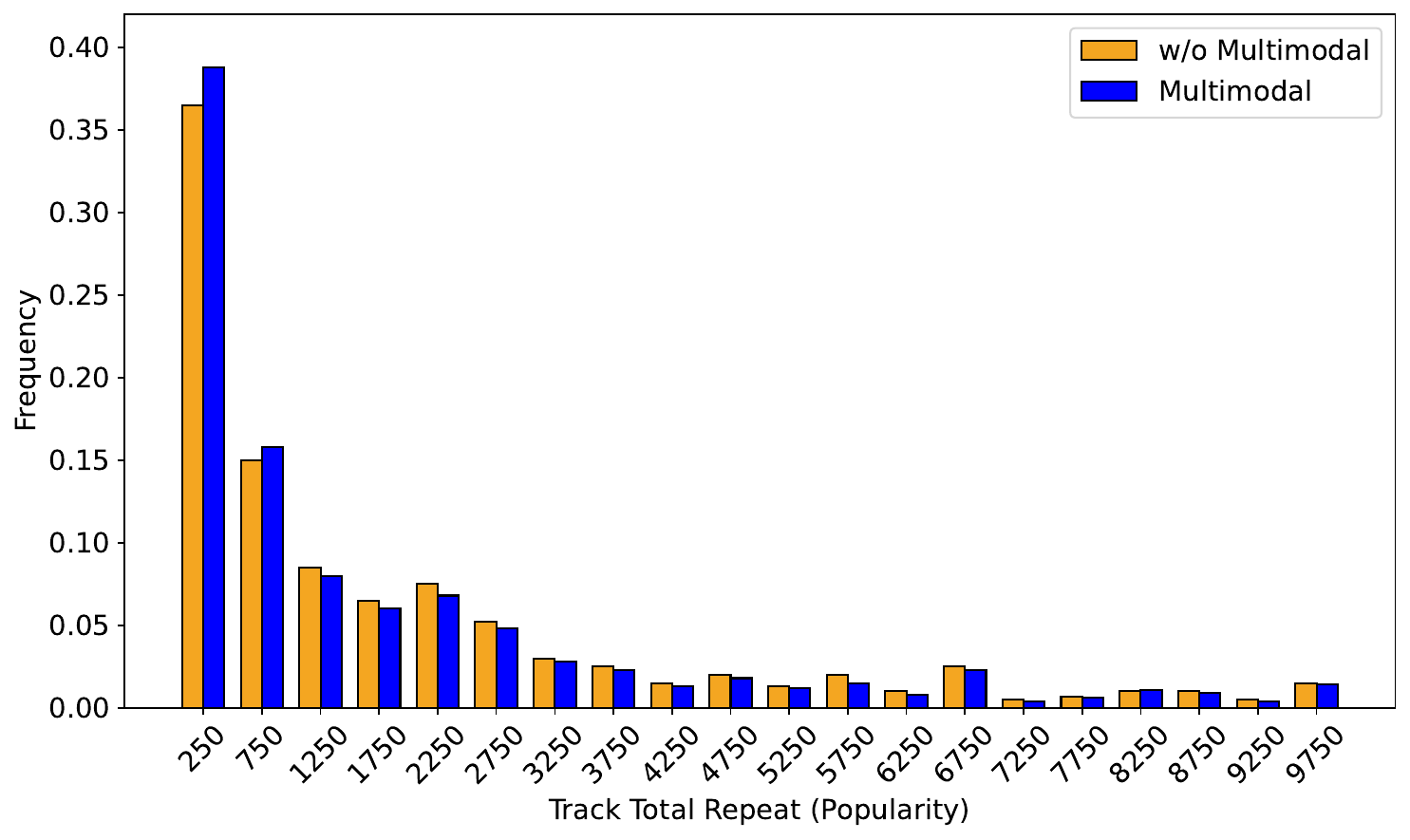}
    \caption{Frequency distribution of Top-10 hits across track popularity bins for multimodal and non-multimodal models, indicating that multimodal features increase the likelihood of recommending less popular tracks.}
    \label{fig:diversity}
\end{figure}

First, we calculate the popularity of all music items by counting their interaction frequency. Then, for each user's candidate items, we score them using our model and select the top-10 predicted items as the model's recommendations. Finally, within a specified range, we observe the popularity distribution of these recommended items. The results are shown in Figure \ref{fig:diversity}. We can see that models utilizing modal features have a higher probability of recommending less popular songs. This indicates that multimodal feature models offer better diversity, enabling users to discover fresher songs.

\subsection{Study on Number of Clusters}

\begin{figure}[h]
    \centering
    \includegraphics[width=0.9\linewidth]{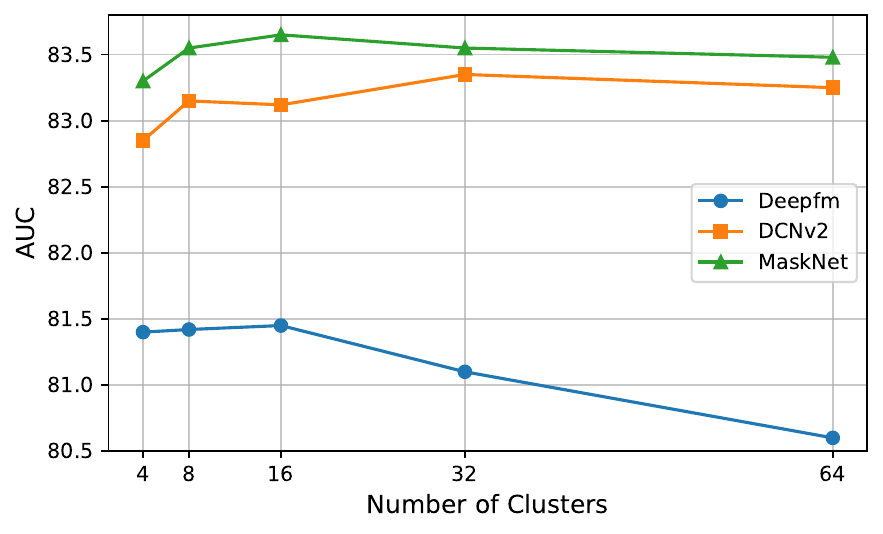}
    \caption{AUC scores under different numbers of clusters.}
    \label{fig:auc_clusters}
\end{figure}

We further experimented with the impact of the number of clusters. The encoder choices are described in subsection \ref{sec:encoders}. The results are shown in figure \ref{fig:auc_clusters}. Clustering effectiveness appears to be maximized when the number of clusters is around 16. 

\subsection{Cold Start Scenarios}

In real-world applications, new songs frequently appear on platforms. These songs lack user interaction history, making their recommendation a critical challenge.\cite{salganik2024larp, lin2024temporally} 
Audio modality is considered crucial for item recommendations in such cases. Therefore, we conduct experiments to verify how multimodal information improves the performance of recommendation systems in item cold-start scenarios. Following previous work \cite{xu2024cmclrec}, we randomly selected a proportion of items as cold-start items and split them 1:1 into the validation and test sets, ensuring they were absent from the training set. The remaining items were divided into training, warm-validation, and warm-testing sets at an 8:1:1 ratio. Next, we created the full validation and test sets by combining their respective warm and cold subsets. The validation set was used for hyper-parameter tuning, and the test set for evaluating performance. We conducted multiple runs under different random seeds. Pairwise two-tailed t-tests were then performed, and these differences were found to be statistically significant (p < 0.05).

We are presenting a selection of representative model results. As shown in Table \ref{tab:cold_m4a} and \ref{tab:cold_2b}, the integration of modal features and our method significantly enhances cold-start effectiveness. As discussed in subsection \ref{sec:diversity}, the Music4all dataset includes item metadata, allowing for some level of prediction for unseen items even without multimodal features. Clearly, however, the addition of multimodal information provides further gains beyond what this metadata offers. In contrast, our LFM-2b-taste setup lacks item metadata input, resulting in poor cold-start performance; the inclusion of multimodal information partially alleviates this issue.
It is worth highlighting that in cold-start scenarios, DCNv2 initially lagged behind AFM when relying solely on metadata. 

However, with the incorporation of multimodal features, DeepFM achieves the highest AUC, demonstrating its improved ability to exploit such information. A similar trend is observed for MaskNet.

\begin{table}[ht]
\centering
\caption{Model Performance Comparison in cold start on Music4all Dataset.}
\resizebox{0.48\textwidth}{!}{
\begin{tabular}{lcc|cc}
\hline
 & \multicolumn{2}{c}{ All } & \multicolumn{2}{c}{ All+MuQ-token }  \\
\hline
\textbf{Model} & \textbf{AUC(\%) $\uparrow$} & \textbf{logloss(\%) $\downarrow$} & \textbf{AUC(\%) $\uparrow$} & \textbf{logloss(\%) $\downarrow$} \\
\hline

EulerNet & 60.67 & 0.4931 & \textbf{61.27} & \textbf{0.4824} \\
FinalMLP & 59.29 & 0.4965 & \textbf{62.09} & \textbf{0.4690} \\
DCNv2 & 62.89 & 0.5235 & \textbf{65.11} & \textbf{0.4840} \\
MaskNet & 59.08 & 0.5154 & \textbf{64.54} & \textbf{0.4767} \\
AFM & 63.84 & 0.4687 & \textbf{64.92} & \textbf{0.4548} \\

\hline
\end{tabular}
}
\label{tab:cold_m4a}
\end{table}

\begin{table}[ht]
\centering
\caption{Model Performance Comparison in cold start on lfm-2b-taste Dataset.}
\resizebox{0.48\textwidth}{!}{
\begin{tabular}{lcc|cc}
\hline
 & \multicolumn{2}{c}{ All } & \multicolumn{2}{c}{ All+MuQ-token }  \\
\hline
\textbf{Model} & \textbf{AUC(\%) $\uparrow$} & \textbf{logloss(\%) $\downarrow$} & \textbf{AUC(\%) $\uparrow$} & \textbf{logloss(\%) $\downarrow$} \\
\hline
EulerNet & 53.82 & 1.0539  & \textbf{60.60} & \textbf{0.9780}  \\
FinalMLP & 50.41 & 0.9099 & \textbf{56.38} & \textbf{0.8140} \\
DCNv2 &  56.00 & 1.4961  & \textbf{59.90}  & \textbf{1.4064} \\
MaskNet & 51.11 & 0.9479 & \textbf{59.17} & \textbf{0.8813} \\
AFM & 52.04 & 0.9047 & \textbf{56.80} & \textbf{0.8233} \\
\hline
\end{tabular}
}
\label{tab:cold_2b}
\end{table}

\section{Conclusion}
This paper introduces the first music recommendation benchmark that leverages audio and text features from self-supervised models. Our primary goal is to address the current lack of modal feature utilization in music recommendation systems.
We built upon three widely used music interaction datasets to create two new datasets enriched with multimodal information. Based on these, we standardized an evaluation framework designed to effectively utilize this information. We conducted extensive experiments across multiple models and settings to validate the performance gains of incorporating audio information into existing classic CTR and recall models. We introduced MuQ-token, an efficient audio integration approach, and demonstrated its efficacy through our benchmark evaluations.

We hope our framework will enable researchers in the music recommendation field to quickly get started and leverage it to advance research in music recommendation systems.

\bibliographystyle{ACM-Reference-Format}
\bibliography{ref}
\clearpage

\section{Appendix}
\appendix

\section{Distribution Drift in Recommendation}

We extracted and observed user and item information for different time periods based on the timestamp information provided by the data. For users, we directly used information such as age, gender, and total number of plays from the dataset. We calculated user features for each time period on a monthly basis, counting the users who interacted during that period. Here, we take the lfm-1b-taste dataset as an example.

 The number of interactions for each user was used as the weight, and the features of all users who interacted were weight-averaged to represent the user features for that time period. The features are then normalized in each field. We statistically track the feature distribution of items over time in a similar manner, using the features extracted by MuQ-mulan. After averaging the features, they are visualized using t-SNE.

\begin{figure}[H]
    \centering
   \includegraphics[width=1\linewidth]{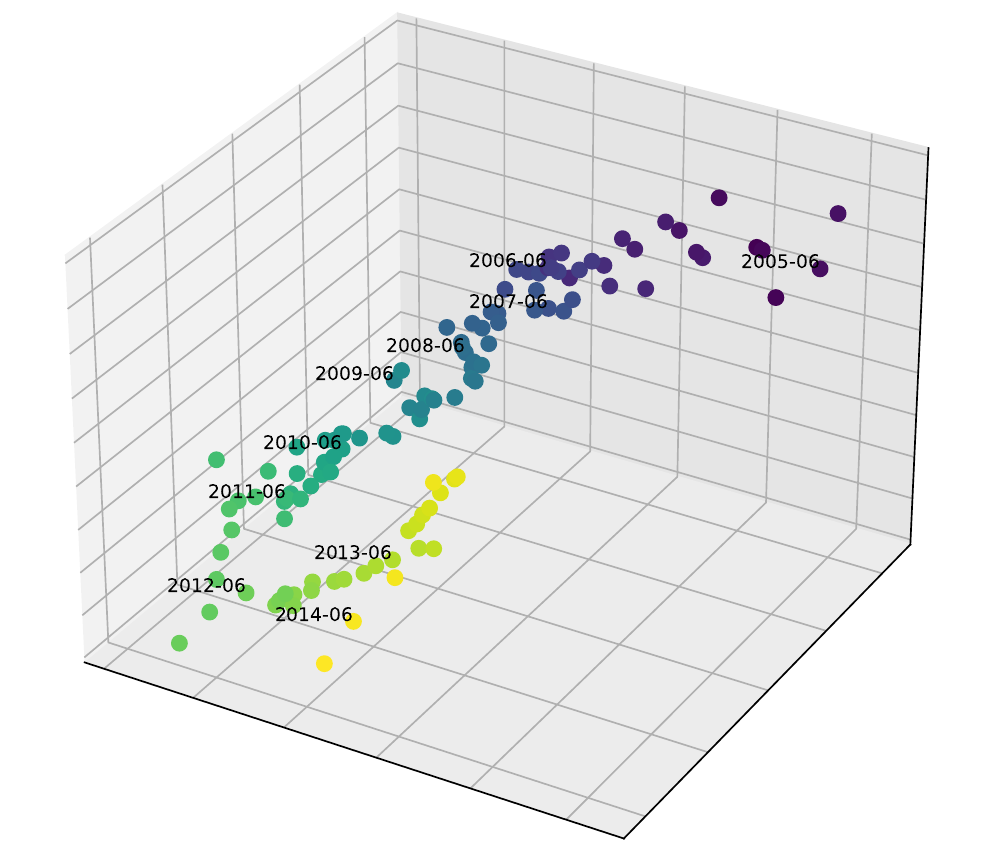}
    \caption{User features distribution across time.}
    \label{fig:user_shift}
\end{figure}

\begin{figure}[H]
    \centering
    \includegraphics[width=\linewidth]{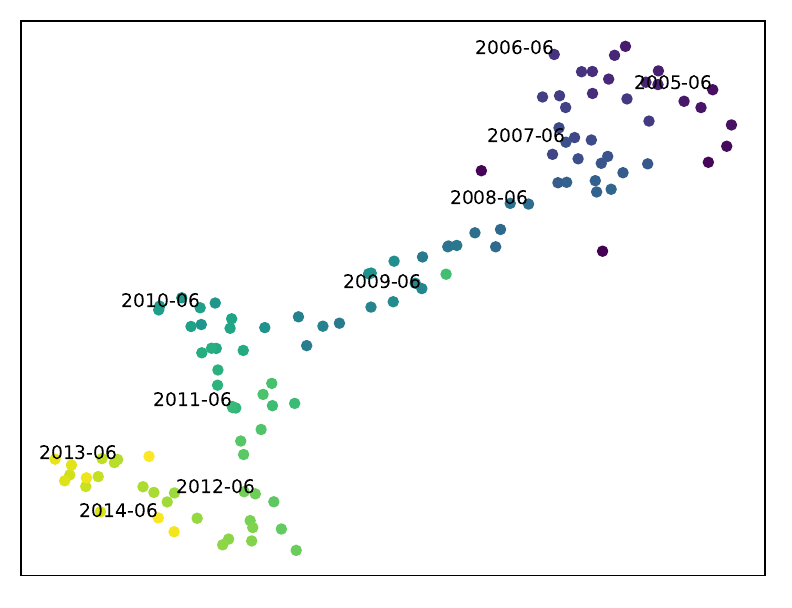}
    \caption{Item features distribution across time.}
    \label{fig:item_shift}
\end{figure}

 As shown in Figures \ref{fig:user_shift} and \ref{fig:item_shift}, we observed that the information distribution of both users and items changes with the temporal shift, which aligns with real-world scenarios.
 As time progresses, these features tend to shift, and during similar time periods, the distribution of features is also relatively close. This issue is a major challenge when deploying models trained on fixed datasets to the real world. This characteristic of the data presents both inspiration and challenges for our research, specifically regarding how to more scientifically evaluate recommendation systems on offline datasets, as mentioned in \cite{wang2023streaming}. Moreover, under this evaluation approach, it raises the question of how to enable models to learn from the changes in data flow, utilizing a small amount of incremental information to quickly adapt to new item and user distributions, thereby performing better in the real world.

\end{document}